# Sensing discomfort of standing passengers in public rail transportation systems using a smart phone

Thommen Karimpanal George, Harit Maganlal Gadhia, Ruben S/O Sukumar
Department of Mechanical Engineering
John-John Cabibihan
Department of Electrical & Computer Engineering
Email: elecjj@nus.edu.sg
National University of Singapore

*Abstract*— **This paper aims to investigate the effect of acceleration on the discomfort of standing passengers. The acceleration levels from different public rail transport lines such as the mass rapid transits (MRTs) and light rail transits (LRTs) of Singapore, as well as the associated qualitative data indicating the discomfort of standing passengers were collected and analyzed. Based on a logistic regression model to analyze the data, a discomfort index was introduced, which can be used to compare various rail lines based on ride comfort. A method for predicting the discomfort of passengers based on the acceleration values was proposed for any given train line.**

## I. Introduction

It is a common sight to see crowded buses and trains in which several passengers remain standing during some part of their journey. The automatic transmission systems used in public transport vehicles produces sudden and high accelerations intermittently throughout the journey. As a result, standing passengers often struggle to maintain balance and experience a significant amount of discomfort at these instances of sudden accelerations and decelerations.

Several studies have been conducted over the years to quantify and relate the passenger comfort levels with the jerks and accelerations of trains. Delton and Dale [1] presented an instrumentation system for the measurement of acceleration profiles associated with mass transit vehicles and maximum allowable jerk and acceleration values for passenger comfort were specified. Ramasamy et al [2] used measured vibrations and conducted questionnaire surveys simultaneously on few long distance trains in Sweden in order to understand the effect of vibration on performing sedentary activities. For this research, ISO-2631 and Sperling Ride Index [3] were used to evaluate the ride comfort. Graaf et al. [4] determined the limiting values for sudden accelerations that cause problems in maintaining postural balance. Omar Bagdadi and András Várhelyi [5] proposed a method for detecting jerks in safety critical events, based on the characteristics of the braking caused by the driver in time critical situations.

In this paper, the accelerations and decelerations in various public rail transportation systems in Singapore and its effects on the discomfort of standing passengers was studied. In order to achieve this objective, the accelerometer of a standard smart phone was used to record the acceleration data from various rail lines. The jerks associated with this acceleration data were analyzed. Simultaneously, the instances of discomfort of standing passengers were recorded while the vehicle was in motion. In this way, a qualitative description of the discomfort of standing passengers was obtained. Based on logistic regression analysis [6] of these data sets, a discomfort index is introduced, which is used to quantify the discomfort experienced during a journey for a particular line. Based on this index, a comparison between the various lines is also presented. Also, using logistic regression techniques, it is shown that for a particular rail line, passenger discomfort can be predicted with reasonable accuracy (typically >80%). Such an analysis can be helpful in a number of applications, including the improvement of current forms of public transport as well as for the design and testing of new transportation systems.

## II. Methodology

The methodology used to conduct this experiment involves recording the acceleration data in different forms of public transportation such as MRTs and LRTs. Accelerometers were placed inside the vehicle at a fixed point such that there was no relative motion between the vehicle and the accelerometer. Owing to the convenience in portability and availability, as well as the ease of use, the accelerometer of a standard smart phone was used to carry out the data collection. In order to ensure quality of the data gathered, in future studies, crowd-sourced intelligence could be made use of. The crowd-sourcing approach to this problem is justified by the fact that smart phones, which are equipped with accelerometers, are becoming more and more prevalent in modern society today [7,8]. Suitable mobile applications such as 'Accelerometer Monitor' (www.mobile-tools.eu), which was used for data collection in this experiment, could act as ideal tools to implement crowd-sourcing, thus leading to an improved data set that would be more representative of the general population.

In this experiment, an android based smart phone with a 3-axis accelerometer was used to measure the accelerations. Of the 3-axes, the Y axis was aligned in the direction of motion of the vehicle, the X axis was at 90° to the Y axis in the same horizontal plane and the Z axis was perpendicular to the plane made by X and Y. The figure below illustrates the directions of accelerations relative to the direction of motion of the train.

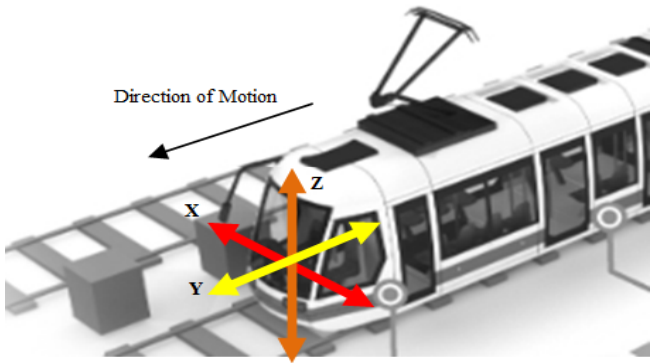

Figure 1. Figure showing the directions of accelerations according to the vehicle motion

Along with the acceleration data, simultaneously, another input, representative of the discomfort, was collected from standing passengers. In order to gather this data, a stopwatch was used so as to record the time instances of passenger discomfort. The standing passenger was instructed to press a stopwatch button which records the time instance corresponding to whenever he/she feels uncomfortable due to the acceleration levels experienced in the train. This set of readings, showing the record of time instances of discomfort was recorded for the entire period of the journey. A plot of the discomfort instances vs time can thus be obtained. Using the sampling rate for the collection of acceleration data, this information was later mapped to the acceleration levels of the train, which was simultaneously recorded using the accelerometer. By comparing these two data sets and plots, the variation in the discomfort of the passenger with the variation in the acceleration levels in the vehicle could be easily visualized and analyzed. Since different passengers have different discomfort tolerances to the same level of acceleration, the experiment was conducted with different standing passengers in order to obtain a data set which is applicable to any standing passenger, in general. For this study, data was collected from three passengers. Although this data set is relatively small, and may not be representative of the general population, the methodology and analysis demonstrated in this paper can be applied to larger data sets as well. As mentioned earlier, for future studies and applications, crowd-sourced intelligence could be made use of in order to obtain more generalized data sets that span a much larger section of the population.

### III. DATA ANALYSIS AND RESULTS

Data collected from the Circle, East-West, North-South and North-East lines of the MRTs as well as from the LRT trains are plotted below. The following plots show the acceleration and jerk values in the three orthogonal directions and the instances of discomfort of the standing passengers. The jerk data is generated by simply differentiating the acceleration data obtained with respect to time. The discomfort instances in these plots are represented with pulse inputs which have non-zero value during periods of discomfort and zero value during all other time instances.

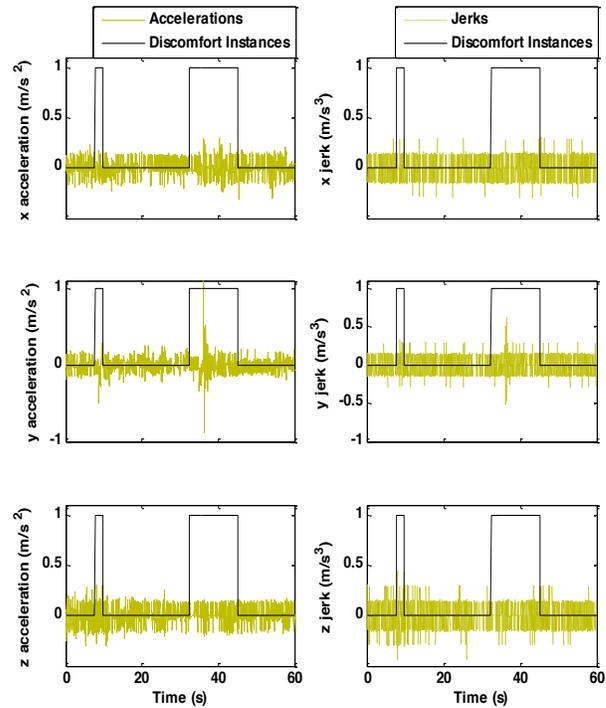

Figure 2. Acceleration and jerk values with discomfort instances plotted against the time for the Circle (Yellow) line.

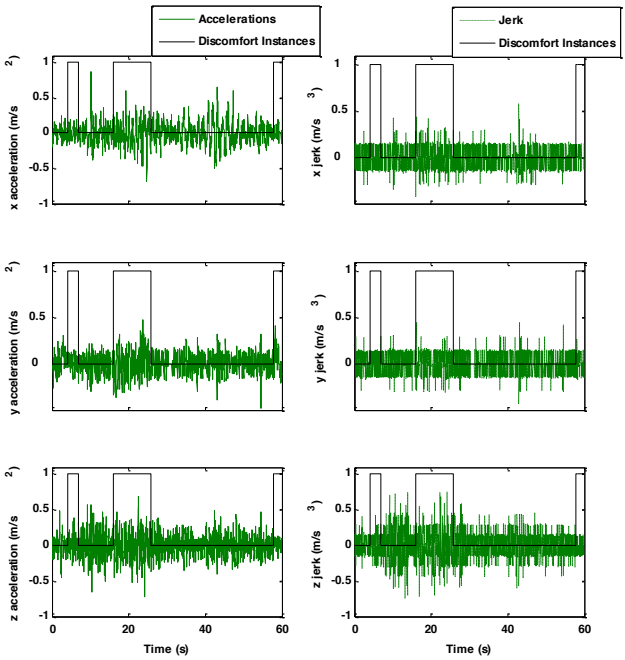

Figure 3. Acceleration and jerk values with discomfort instances plotted against time for the East-West (Green) line.

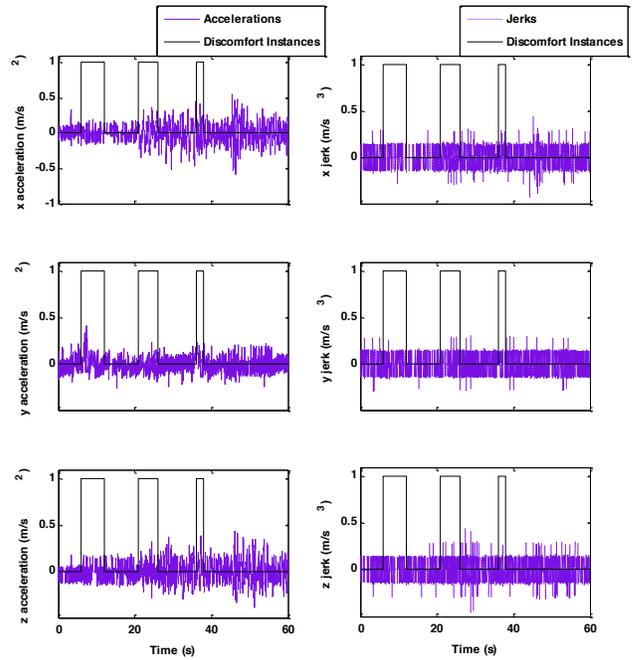

Figure 5. Acceleration and jerk values with discomfort instances plotted against time for the North-East (purple) line.

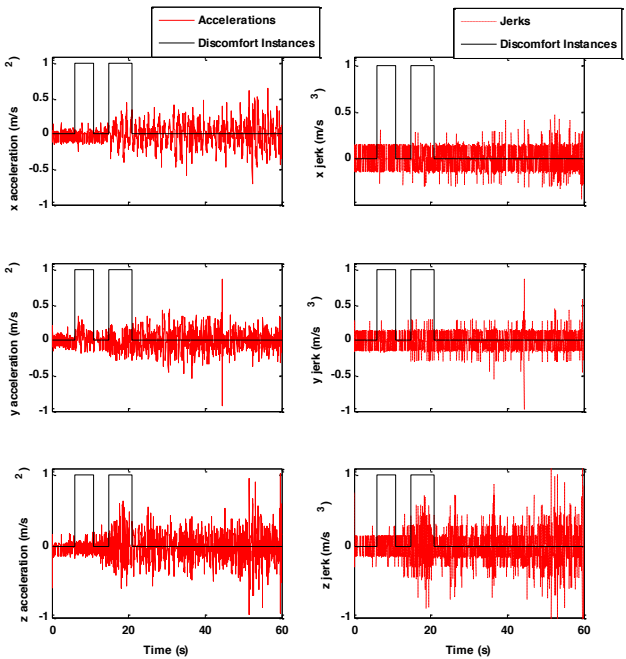

Figure 4. Acceleration and jerk values with discomfort instances plotted against time for the North-South(red) line.

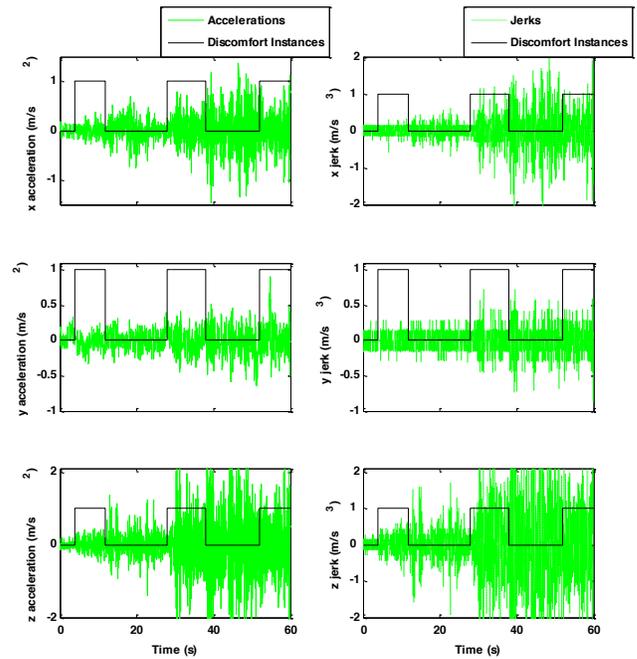

Figure 6. Acceleration and jerk values with discomfort instances plotted against time for LRT

It is clear from the graphs shown, that just by visual inspection; no intuitive relation can be deduced between the acceleration values and the instances of discomfort. The discomfort caused may be attributed to a single component of acceleration or a combination for two or more acceleration components. Similarly, no clear relation can be deduced between the jerk values and the instances of discomfort. Thus, analysis of the data is done in order to establish some form of relationship between the two quantities. Those instances where the passengers recorded discomfort were labeled '1' and all other instances were labeled '0'. As there are only two cases of output, that is, comfortable or uncomfortable, this is a problem of binary classification. Hence, the technique of logistic regression is used to analyze this set of binary data [5]. Other classification methods can also be used. In logistic regression, the probability of obtaining a positive case is given by [7]:

$$P = \frac{1}{1+e^{-z}} \quad (1)$$

Where 'P' is the probability and 'z' is a linear function in the features used. In this case, the relevant features are the acceleration values along the three orthogonal axes. By calculating the value of probability for a given set of acceleration values in a particular line, it is shown that the state of comfort or discomfort of the passengers can be predicted.

The above equation indicates that the task of classification into either case is dependent on the parameter 'z'. In terms of the three acceleration values, the value of 'z' is given as:

$$z = a_x b_1 + a_y b_2 + a_z b_3 + b_4 \quad (2)$$

Thus, the value of 'P' is ultimately decided by the constants $b_1, b_2, b_3$ and $b_4$ and the normalized values of the acceleration components along the three axes $a_x, a_y$ and $a_z$. Here, the coefficients are determined by using logistic regression analysis. In this experiment, the coefficients are solved for using MATLAB. The coefficients in 'z' are characteristic of a particular line. Hence, each line has a statistical equation associated with it, which can be used to predict the level of discomfort, given the values of acceleration. The equations for the lines investigated in this paper are shown below in the Table I:

TABLE I

| Line | Discomfort Equation |
|---|---|
| Circle (yellow) Line | $z = a_x(0.1728) + a_y(1.2064) + a_z(-0.9458) + (-1.1528)$ |
| North-South(red) line | $z = a_x(-1.1764) + a_y(-0.2396) + a_z(-1.0849) + (-3.283)$ |
| North-East (purple) line | $z = a_x(0.8223) + a_y(0.2607) + a_z(-0.7349) + (-0.9995)$ |
| East-West (green) line | $z = a_x(-0.2867) + a_y(3.3714) + a_z(0.4978) + (1.9170)$ |
| LRT | $z = a_x(-0.7117) + a_y(0.4723) + a_z(-0.2185) + (-1.2651)$ |

Discomfort equations for each line *(Table I)*

Observing the values associated with each acceleration component reveals how much the discomfort experienced is dependent on that component of acceleration. For example, in the equation for the East-West (green) line, since the value associated with the y component of acceleration is significantly higher than those associated with the other two components, we can deduce that the discomfort experienced on the East-West line is mainly contributed by the acceleration values in the y direction. Similar analysis can be carried out for other lines. Such information could be useful to maintenance engineers who could take suitable, directed action to improve the ride comfort. Equations (1) and (2) can also be used to predict whether a given set of acceleration data for a particular rail line is comfortable or uncomfortable for a passenger, in general. The prediction for a test data set of the East-West line is shown in figure 7:

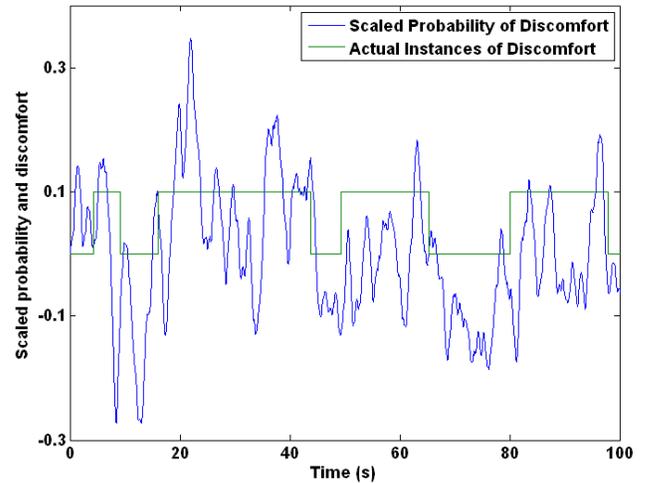

Figure 7. Figure showing predicted and actual instances of discomfort for a test data set of the East-West (Green) line

The figure above shows the plot of predicted state of discomfort for a test data set of about 100 seconds (5000 samples) for the East-West (green) line. The state of discomfort is predicted using the probability function in (1). In this figure, the probability is filtered and scaled for better viewing. The plots show that whenever the scaled probability curve rises above '0', a state of discomfort is predicted. The prediction is validated by the actual discomfort pulse which was recorded for that particular journey, which is also shown in the plot. Such prediction can be useful to train manufacturers in the testing stage, in order to predict whether passengers will feel comfortable with the levels of acceleration generated by the trains.

As another outcome of the analysis, a discomfort index D, based on equation (1) and a modified version of equation (2) is introduced. The discomfort index is given by:

$$D = \frac{1}{1+e^{-Z}} \quad (3)$$

Here, the value of Z is calculated as follows:

$$Z = A_x b_1 + A_y b_2 + A_z b_3 + b_4 \quad (4)$$

Where $A_x$, $A_y$ and $A_z$ are the mean values of the normalized acceleration components along the three axes in the training data set.

Hence, D gives an overall discomfort value for a particular line. The higher the value of D, the greater is the discomfort experienced. Based on this index, a comparison of the different lines in terms of standing passenger comfort is depicted in figure 13:

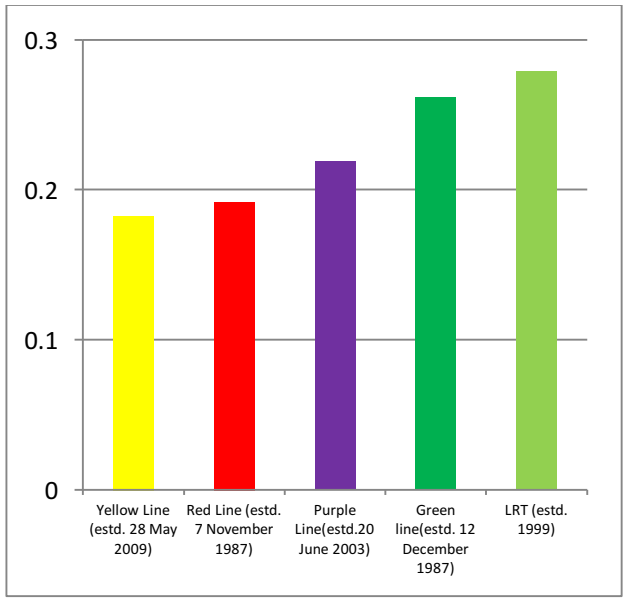

Figure 8. Figure showing the discomfort indexes for different rail lines

This comparison can be extended to other forms of public transport as well, where passengers are often required to remain standing. Such an index may act as a guideline for design engineers while designing new public transport systems. This index may also be useful in evaluating road surfaces. In public buses, the discomfort index could be used to evaluate the driving style of the bus driver. However, for this, real time traffic conditions, road surface conditions and other variables should also be taken into consideration.

Alternatively, this analysis may also be carried out using the jerk data. The data-set for jerk can be obtained by differentiating the acceleration values with respect to time. The later part of the analysis will remain similar to that using acceleration values.

IV. CONCLUSION

Accelerometer and discomfort data were collected and analyzed using the logistic regression technique. Discomfort equations were defined for each of the different rail lines. Using these equations, it is shown that for a particular line, the discomfort of standing passengers can be predicted, given the acceleration values. Further, a discomfort index is introduced, which can be used to compare different modes of transport on the basis of ride comfort for standing passengers. Other applications of this index, such as evaluating road surfaces and driving styles have also been briefly discussed.